\documentclass[preprint,prd,nofootinbib,tightenlines,groupedaddress,amsmath,amssymb]{revtex4}

\usepackage{bm}
\usepackage{color}
\usepackage{graphicx}
\allowdisplaybreaks
\usepackage{multirow}

\begin{document}


\title{
CP Violating Polarization Asymmetry in Charmless \\Two-Body  Decays of Beauty Baryons}

\author{Min He$^{1}$\footnote{hemin\_sjtu@163.com}}
\author{Xiao-Gang He$^{1,2,3}$\footnote{hexg@phys.ntu.edu.tw}}
\author{Guan-Nan Li${}^{4,2}$\footnote{lgn198741@126.com}}

\affiliation{${}^{1}$INPAC, SKLPPC and Department of Physics and Astronomy,
Shanghai Jiao Tong University, Shanghai, China}
\affiliation{${}^{2}$CTS, CASTS and
Department of Physics, National Taiwan University, Taipei, Taiwan}
\affiliation{${}^{3}$Physics Division, National Center for
Theoretical Sciences, Hsinchu, Taiwan}
\affiliation{${}^{4}$Department of Physics, Zhengzhou University, Zhengzhou, Henan, China}

\date{\today $\vphantom{\bigg|_{\bigg|}^|}$}

\date{\today}

\vskip 1cm
\begin{abstract}
Several baryons containing a heavy b-quark, the b-baryons, have been discovered.  The charmless two-body decays of b-baryons can provide a new platform for CP violating studies in a similar way as charmless two-body decays of B-meson. In b-baryon decays there are new CP violating observable related to baryon polarization. We show that in the flavor $SU(3)$ limit there exist relations involve different combinations of the decay amplitudes compared with those in CP violating rate asymmetry. These new relations therefore provide interesting tests for the mechanism of CP in the standard model (SM) and flavor $SU(3)$ symmetry. Future data from LHCb can test these relations.
\end{abstract}

\pacs{PACS numbers: }

\maketitle


\noindent
{\bf Introduction}

Several baryons containing a heavy b-quark, the beauty baryon (b-baryon) $\cal B$,  have been discovered~\cite{pdg}.  The study of heavy mesons containing a b-quark, the $B$ mesons, provided crucial information~\cite{pdg} in establishing the standard model (SM) for $CP$ violation, the Cabibbo-Kobayashi-Maskawa (CKM) model~\cite{km}. The decays of the $\cal B$ b-baryons can provide a new platform to further test the CKM model of CP violation~\cite{cdf1,model,gronau,rateasymmetry}.
It has been shown that  ${\cal B}$ b-baryon decay into a light $SU(3)$ octet baryon ${\cal F}$ and a light $SU(3)$ octet pseudoscalar meson ${\cal M}$, the charmless two-body b-baryon decay ${\cal B} \to {\cal M} + {\cal F}$,  can have sizeable CP violation\cite{model}.  There are also flavor $SU(3)$\cite{su3} relations between CP violating rate asymmetries in some of these decays\cite{rateasymmetry}. 
Similar relations for CP violating rate asymmetry have been obtained for $B$ meson decays before for two-body charmless $B$ meson decays~\cite{he1,he2,he3,he4}. Experimental data have verified some of the relations to very good precision\cite{pdg,rateasymmetry,he4,exp-relation}. 
Therefore it is interesting to see whether relations based on $SU(3)$ symmetry are accidental or are more universal for charmless two-body decays of hadrons containing a heavy b-quark. Experimental data from LHCb can provide crucial information when CP violation for two-body charmless b-baryon decays will be measured. In this paper we study further possible new CP violating observables in $\cal B \to \cal M + \cal F$ decays associated with the polarization of baryon in the decays. We find that there are relations exist in CP violating polarization asymmetries in some of the decays channel and can be tested at the LHCb.
\\

\noindent
{\bf $CP$ Violating Asymmetry for  b-baryon Decays}

The effective Hamiltonian inducing
$\cal B \to \cal M + \cal F$ decays in the SM has both parity (P) conserving and violating parts given by
\begin{eqnarray}
 H_{eff}(q) = {4 G_{F} \over \sqrt{2}} [V_{ub}V^{*}_{uq} (c_1 O_1 +c_2 O_2)
   - \sum_{i=3}^{12}(V_{ub}V^{*}_{uq}c_{i}^{uc} +V_{tb}V_{tq}^*
   c_i^{tc})O_{i}],
\end{eqnarray}
where $q$ can be $d$ or $s$. $V_{ij}$ is the CKM matrix element. In the above the factor $V_{cb}V_{cq}^*$ has
been eliminated using the unitarity property of the CKM matrix.The coefficients
$c_{1,2}$ and $c_i^{jk}=c_i^j-c_i^k$, with $j$ and $k$ indicate the internal quark,
are the Wilson Coefficients (WC) which WC have been studied by several groups and can be
found in Ref.~\cite{heff}. The operators $O_i$ are
 given by
\begin{eqnarray}
\begin{array}{ll}
O_1=(\bar q_i u_j)_{V-A}(\bar u_i b_j)_{V-A}\;, &
O_2=(\bar q u)_{V-A}(\bar u b)_{V-A}\;,\\
O_{3,5}=(\bar q b)_{V-A} \sum _{q'} (\bar q' q')_{V \mp A}\;,&
O_{4,6}=(\bar q_i b_j)_{V-A} \sum _{q'} (\bar q'_j q'_i)_{V \mp A}\;,\\
O_{7,9}={ 3 \over 2} (\bar q b)_{V-A} \sum _{q'} e_{q'} (\bar q' q')_{V \pm A}\;,\hspace{0.3in} &
O_{8,10}={ 3 \over 2} (\bar q_i b_j)_{V-A} \sum _{q'} e_{q'} (\bar q'_j q'_i)_{V \pm A}\;,\\
O_{11}={g_s\over 16\pi^2}\bar q \sigma_{\mu\nu} G^{\mu\nu} (1+\gamma_5)b\;,&
O_{12}={Q_b e\over 16\pi^2}\bar q \sigma_{\mu\nu} F^{\mu\nu} (1+\gamma_5)b\;,
\end{array}
\end{eqnarray}
where $(\bar a b)_{V\pm A} = \bar a \gamma_\mu (1\pm \gamma_5) b$, $G^{\mu\nu}$ and
$F^{\mu\nu}$ are the field strengths of the gluon and photon, respectively. $O_{1,2}$ , $O_{3, 4,5,6}$ and $O_{7,8,9,10}$ are the tree, penguin and electroweak penguin operators.
$O_{11,12}$ are the photonic and gluonic dipole penguin operators

The above effective Hamiltonian dictates that the b-baryon decay amplitude $\cal A$ to have both parity conserving $A_c$ and violating $A_v$ amplitudes in the form
\begin{eqnarray}
{\cal A} = \bar {\cal F} (A_v + iA_c \gamma_5) {\cal B} = {\cal S} +{\cal P} \sigma\cdot \vec p_c\;.
\end{eqnarray}
Here $\vert \vec p_{c}\vert =\sqrt{E^2_{\cal F}-m^2_{\cal F}}$ is size of the final baryon ${\cal F}$ momentum $p_c$.        $m_{\cal B}$ and $m_{\cal M}, m_{\cal F}$ are the masses of the initial baryon and the final meson and baryon particles.
$E_{\cal F}$ is the energy of the final baryon.

The amplitudes ${\cal S}$ and ${\cal P}$ are the  so called $S$-wave (parity = $-$) and $P$-wave (parity = $+$) amplitudes respectively.  Their relations to $A_{v}$ and $A_{c}$ are given by
\begin{eqnarray}
{\cal S}=A_v \sqrt{{(m_{\cal B}+m_{\cal F})^2-m^2_{\cal M} \over  16 \pi m^2_{\cal B}}}\;,\;\;{\cal P}= A_c\sqrt{{(m_{\cal B}-m_{\cal F})^2-m^2_{\cal M} \over 16\pi m^2_{\cal B}}}\;.
\end{eqnarray}

The anti-b-baryon decay amplitude $\bar {\cal A}$ is usually written as
\begin{eqnarray}
\bar {\cal A} = - \bar {\cal S} +\bar {\cal P} \sigma\cdot \vec p_c\;.
\end{eqnarray}
In the limit of CP conservation, $\bar {\cal S} = {\cal S}$ and $\bar {\cal P} = {\cal P}$.

In the SM, there are tree and penguin contributions to  ${\cal S}$ and ${\cal P}$ for $\Delta S=0$ ($q=d$) and $\Delta S=-1$ ($q=s$).
The ${\cal S}$ and ${\cal P}$ and their corresponding anti-particle decay amplitudes can be decomposed into tree $T_i$ and penguin $P_i$ amplitudes as
\begin{eqnarray}
&&{\cal S}(q)=V_{ub}V^{*}_{uq}T(q)_{0}+V_{tb}V^{*}_{tq}P(q)_{0}\;,\;\;
{\cal P}(q)=V_{ub}V^{*}_{uq}T(q)_{1}+V_{tb}V^{*}_{tq}P(q)_{1}\;,\nonumber\\
&&\bar{{\cal S}}(q)=V^{*}_{ub}V_{uq}T(q)_{0}+V^{*}_{tb}V_{tq}P(q)_{0}\;,\;\;
\bar{{\cal P}}(q)=V^{*}_{ub}V_{uq}T(q)_{1}+V^{*}_{tb}V_{tq}P(q)_{1}\;,
\end{eqnarray}
where the sub-indices $0,1$  denote the $S$ and $P$ amplitudes.

The decay width  is given by
\begin{eqnarray}
\Gamma&=&2 p_{c}  (\vert{\cal S}\vert^2+\vert {\cal P}\vert^2)\;,
\end{eqnarray}

If information on the polarization of the decay is available, there are additional experimental observables  in the decay angular distribution.
In the rest frame of the initial b-baryon, the angular distribution can be expressed as\cite{hyperon}
\begin{eqnarray}
{4\pi \over \Gamma} {d \Gamma \over d \Omega} =1+\alpha \vec{s}_{\cal B}\cdot\vec{n}
+\vec{s}_{\cal F}\cdot[(\alpha+\vec{s}_{\cal B}\cdot\vec{n})\vec{n}
+\beta\vec{s}_{\cal B}\times\vec{n} +\gamma(\vec{n}\times(\vec{s}_{\cal B}\times\vec{n}))]\;,
\end{eqnarray}
where $\vec{s}_{\cal B},\vec{s}_{\cal F}$ are the spins of initial b-baryon and final octet baryon, and $\vec{n}=
\vec{p_{c}}/\vert p_{c}\vert$ is the direction of the final baryon $\cal F$. The parameters $\alpha$, $\beta$ and $\gamma$ expressed in terms of the ${\cal S}$ and ${\cal P} $ amplitudes are given by
\begin{eqnarray}
\alpha={{2Re({\cal S}^{*}{\cal P})}\over {\vert{\cal S}\vert^2+\vert{\cal P}\vert^2}},\;
\beta={{2Im({\cal S}^{*}{\cal P})}\over {\vert{\cal S}\vert^2+\vert{\cal P}\vert^2}},\;
\gamma={ {\vert{\cal S}\vert^2-\vert{\cal P}\vert^2}  \over   {\vert{\cal S}\vert^2+\vert{\cal P}\vert^2}}
\end{eqnarray}
only two of them are independent with $\alpha^2+\beta^2+\gamma^2=1$. With this constraint one can obtain
\begin{eqnarray}
\beta=(1-\alpha^2)^{{1\over2}}sin{\phi},\; \gamma=(1-\alpha^2)^{{1\over2}}cos{\phi},\; \phi=tan^{-1}(\beta/\gamma)
\end{eqnarray}

Experimentally, to obtain $\alpha$ one needs to measure initial baryon or final baryon polarization. The initial polarization may be difficult to obtain in the LHCb experiment. But the final baryon polarization may be measured by looking at the final baryon decays\cite{valencia}. To obtain $\beta$ and $\gamma$, one needs to have both the initial and final baryon polarization information.  In the rest of the discussions, we will only discuss properties related to $\alpha$ which has the chance to be measured.

For $CP$ violation studies, one can define new quantities related to polarization parameters which vanish in the CP conservation limit. To this end, we study the particle and anti-particle polarization difference as a new CP violating observable
\begin{eqnarray}
\Delta_\alpha(q) &=& \Gamma(q) \alpha(q) + \bar{\Gamma}( q)\bar \alpha( q)\;.
\end{eqnarray}

Express $\Delta_\alpha$ in terms of the ${\cal S}$ and ${\cal P}$ amplitudes, we have
\begin{eqnarray}
\Delta_\alpha &=& 4p_c (Re({\cal S}^{*}{\cal P})-Re({\bar{\cal S}}^{*} {\bar{\cal P}}))\;,\nonumber\\
&=& p_c Im(V_{ub}V^{*}_{uq}V^{*}_{tb}V_{tq})[Im(T(q)^{*}_{0}P(q)_{1})-Im(P(q)^{*}_{0}T(q)_{1})]\;.
\end{eqnarray}

In the flavor $SU(3)$ limit, $U$-spin ($d$ and $s$ exchange) related decay modes have the same tree and penguin amplitudes, that is $T(d)_{j}=T(s)_{j}$ and $P(d)_{j}=P(s)_{j}$. Using the well known relation of the CKM matrix, $Im(V_{ub}V^{*}_{ud}V^{*}_{tb}V_{td})=-Im(V_{ub}V^{*}_{us}V^{*}_{tb}V_{ts})$,  one therefore has
\begin{eqnarray}
\Delta_\alpha(d) = -\Delta_\alpha(s)\;.
\end{eqnarray}

For the weighted polarization asymmetry, $
{\cal A}_\alpha=({\Gamma\alpha +\bar{\Gamma}\bar{\alpha}})/({\Gamma +\bar{\Gamma}})$,
we have
\begin{eqnarray}
&&{{\cal A}_\alpha ({\cal B}_a \to {\cal M} {\cal F} )_{\Delta S = 0}\over {\cal A}_{\alpha}({\cal B}_b \to  {\cal M} {\cal F})_{\Delta S = -1}}
=
- {Br({\cal B}_b \to {\cal M} {\cal F} )_{\Delta S = -1}\over Br({\cal B}_a \to  {\cal M} {\cal F})_{\Delta S = 0}}\cdot
{{\tau_{{\cal B}_a} } \over  {\tau_{{\cal B}_b }} }\;,\label{poasy}
\end{eqnarray}
where $\tau_{{\cal B}_{a,b}}$ indicate the lifetimes of b-baryons ${\cal B}_{a,b}$, and $Br$ indicates branching ratio.

Let us compare a $U$-spin  relation for rate asymmetry obtained in Ref.\cite{rateasymmetry},
\begin{eqnarray}
&&{A_{CP}({\cal B}_a \to {\cal M} {\cal F} )_{\Delta S = 0}\over A_{CP}({\cal B}_b \to  {\cal M} {\cal F})_{\Delta S = -1}}
=
- {Br({\cal B}_b \to {\cal M} {\cal F} )_{\Delta S = -1}\over Br({\cal B}_a \to  {\cal M} {\cal F})_{\Delta S = 0}}\cdot
{{\tau_{{\cal B}_a} } \over  {\tau_{{\cal B}_b }} }\;, \label{raasy}
\end{eqnarray}
where $A_{CP} (q)= \Delta(q)/(\Gamma(q)+\bar{\Gamma}( q))$ with
\begin{eqnarray}
\Delta(q)&=&\Gamma(q)-\bar{\Gamma}(q) = 2 p_c \left ( |{\cal S}|^2+|{\cal P}|^2 - (|\bar {\cal S}|^2+|\bar {\cal P}|^2)\right )\;,\nonumber\\
&=&-8 p_{c} Im(V_{ub}V^{*}_{uq}V^{*}_{tb}V_{tq})Im[T(q)_{0}P(q)^{*}_{0}+T(q)_{1}P(q)^{*}_{1}]\;,
\end{eqnarray}
and in the $SU(3)$ limit, the $U$-spin related states have
\begin{eqnarray}
\Delta(d)&=&-\Delta(s)\;.
\end{eqnarray}

It is clear that the relations in  eq.(\ref{poasy}) and eq.(\ref{raasy}) test different aspects of $CP$
violating properties in the decay amplitudes. The rate asymmetry $A_{CP}$ probes $CP$ violation due to interference between tree and penguin amplitudes of the same parity, while the polarization asymmetry ${\cal A}_\alpha$ probes $CP$ violation due to interference between tree and penguin amplitudes of opposite parities. The relation associated with rate asymmetry has been discussed. The relation associated with the polarization asymmetry is new and is our emphasis in this paper.

To test the above relations, one needs to make sure that indeed there are processes which are related by $U$-spin as described as the above. We show next that indeed such relations exist in two-body charmless decays of both anti-triplet and sextet, and tests for some of the relations can be carried out experimentally.
\\

\noindent
{\bf Amplitude relations in Two-body Chamrless b-Baryon Decays}

We now study the charmless two-body decays decay amplitudes of low-lying b-baryon. The low-lying ${1\over 2}^+$ $\cal B$ b-baryons are made up of a $b$ quark and two light quarks. Here the light quark $q$ is one of the $u$, $d$ or $s$ quarks.
Under the flavor $SU(3)$ symmetry, the $b$ quark is a singlet and the light quark $q$ is a member in the fundamental representation $3$.  The b-baryons then have representations under flavor $SU(3)$ as $1\times 3 \times 3 \time 3 = \bar 3 + 6$, that is,  the b-baryons contain an anti-triplet and a sextet in the  $SU(3)$ flavor space\cite{heavy-baryon}.  The anti-triplet $\cal B$ and the sextet ${\cal C}$ b-baryons will be indicated by
 \begin{eqnarray}
 ({\cal B}_{ij})=\left ( \begin{array}{ccc}0&\Lambda^{0}_{b}&\Xi^{0}_{b}\\
 -\Lambda^{0}_{b}&0&\Xi^{-}_{b}\\
 -\Xi^{0}_{b}&-\Xi^{-}_{b}&0
 \end{array}
\right )\;,\;\;{\cal C}_{ij}=\left ( \begin{array}{ccc}
\Sigma^{+}_{b}& {\Sigma^{0}_{b}\over \sqrt{2}} &{\Xi'^{0}_{b}\over \sqrt{2}}\\
{\Sigma^{0}_{b}\over \sqrt{2}} & \Sigma^{-}_{b} &{\Xi'^{-}_{b}\over \sqrt{2}}\\
{\Xi'^{0}_{b}\over \sqrt{2}} &  {\Xi'^{-}_{b}\over \sqrt{2}} &\Omega^{-}_{b}
\end{array}
\right )\;.
 \end{eqnarray}
Their quark compositions are as the following\cite{heavy-baryon}
\begin{eqnarray}
&&\Lambda^{0}_{b}={1\over\sqrt{2}}(ud-du)b\;,\;\; \Xi^{0}_{b}={1\over\sqrt{2}}(us-su)b\;,\;\; \Xi^{-}_{b}={1\over\sqrt{2}}(ds-sd)b\;,\nonumber\\
&&\Sigma^{+}_{b}=uub\;,\;\;\Sigma^{0}_{b}={1\over \sqrt{2}}(ud+du)b\;,\;\; \Sigma^{-}_{b}=ddb\;, \nonumber\\
&&\Xi'^{0}_{b}={1\over \sqrt{2}}(us+su)b\;,\;\;\Xi'^{-}_{b}={1\over \sqrt{2}}(ds+sd)b\;,\;\;\Omega^{-}_{b}=ssb\;.
\end{eqnarray}

The two charmless states in the final state of $\cal B$ decay are the ${1\over 2}^+$  octet baryons $\cal F$ and the pseudoscalar octet mesons
$\cal M$, respectively. They are
\begin{eqnarray}
({\cal M}_{ij}) = \left ( \begin{array}{ccc}
{\pi^{0}\over \sqrt{2}}+{\eta_{8} \over \sqrt{6}}& \pi^{+}& K^{+}\\
\pi^{-}& -{\pi^{0}\over \sqrt{2}}+{\eta_{8} \over \sqrt{6}}&{K}^{0}\\
K^{-}&  \bar K^{0} & -{2\eta_{8} \over \sqrt{6}}
\end{array}
\right )\;,\;\;
({\cal F}_{ij}) = \left ( \begin{array}{ccc}
{\Sigma^{0}\over \sqrt{2}}+{\Lambda^{0} \over \sqrt{6}}& \Sigma^{+}& p\\
\Sigma^{-}& -{\Sigma^{0}\over \sqrt{2}}+{\Lambda^{0} \over \sqrt{6}}&n\\
\Xi^{-}&  \Xi^{0} & -{2\Lambda^{0} \over \sqrt{6}}
\end{array}
\right )\;.
\end{eqnarray}

At the hadron level, the decay amplitude can be generically written as
\begin{eqnarray}
{\cal A} = \langle {\cal F} {\cal M}\vert H_{eff}(q)\vert {\cal {B}}\rangle =  V_{ub}V^*_{uq} T(q) + V_{tb}V^*_{tq}P(q).
\end{eqnarray}

The operators $O_{i}$ contains $ \overline{3}$, $6$ , $\overline{15}$ of flavor $SU(3)$ irreducible representations. Indicating these representations by matrices $ H(\overline{3})$,
$H(6)$, $H(\overline{15})$~\cite{su3,he1}.
The non-zero entries of the matrices $H(i)$ are given as the followed~\cite{su3, he1}.

For $\Delta S=0$,
\begin{eqnarray}
&&H(\overline{3})^{2}=1\;,\; H(6)^{12}_{1}=H(6)^{23}_{3}=1\;,
\;H(6)^{21}_{1}=H(6)^{32}_{3}=-1\;,\nonumber\\
&&H(\overline{15})^{12}_{1}=H(\overline{15})^{21}_{1}=3,\;H(\overline{15})^{22}_{2}=-2\;,
\;H(\overline{15})^{32}_{3}=H(\overline{15})^{23}_{3}=-1\;,
\end{eqnarray}
and for $\Delta S=-1$,
\begin{eqnarray}
&&H(\overline{3})^{3}=1\;,\; H(6)^{13}_{1}=H(6)^{32}_{2}=1\;,
\;H(6)^{31}_{1}=H(6)^{23}_{2}=-1\;,\nonumber\\
&&H(\overline{15})^{13}_{1}=H(\overline{15})^{31}_{1}=3\;,\;H(\overline{15})^{33}_{3}=-2\;,
\;H(\overline{15})^{32}_{2}=H(\overline{15})^{23}_{2}=-1\;.
\end{eqnarray}

To obtain the $SU(3)$ invariant decay amplitude for a b-baryon, one first uses the Hamiltonian to annihilate the b-quark in ${\cal B}$ and then contract $SU(3)$ indices in an appropriate way with final states $\cal F$ and $\cal M$. As far as $SU(3)$ properties are concerned, the $\cal S$ and $\cal P$ amplitudes will have various $SU(3)$ irreducible amplitudes. Taking the anti-triplet tree amplitude $T_t(q)_0$ and sextet tree amplitude $T_s(q)_0$ in $\cal S$ as examples, we have\cite{rateasymmetry}

\begin{eqnarray}
T_{t}(q)_0&=&a_t(\overline{3})\langle {\cal F}^{k}_{l}{\cal M}^{l}_{k}\vert H(\overline{3})^{i} \vert {\cal B}_{i'i''}\rangle \epsilon ^{ii'i''}
+b_t(\overline{3})_{1}\langle {\cal F}^{k}_{j} {\cal M}^{i}_{k} \vert H(\overline{3})^{j}\vert {\cal B}_{i'i''}\rangle \epsilon ^{ii'i''}\nonumber\\
&+&b_t(\overline{3})_{2}\langle {\cal F}^{i}_{k}{\cal M}^{k}_{j}\vert H(\overline{3})^{j} \vert {\cal B}_{i'i''}\rangle \epsilon ^{ii'i''}
+a_t(6)_{1}\langle {\cal F}^{k}_{l} {\cal M}^{l}_{j} \vert H(6)^{ij}_{k}\vert {\cal B}_{i'i''}\rangle \epsilon ^{ii'i''}\nonumber\\ &+&a_t(6)_{2}\langle {\cal F}^{l}_{j}{\cal M}^{k}_{l}\vert H(6)^{ij}_{k}\vert {\cal B}_{i'i''}\rangle \epsilon ^{ii'i''}
+b_t(6)_{1}\langle {\cal F}^{l}_{k} {\cal M}^{i}_{j}\vert H(6)^{jk}_{l}\vert {\cal B}_{i'i''}\rangle \epsilon ^{ii'i''}\nonumber\\
&+&b_t(6)_{2}\langle {\cal F}^{i}_{j}{\cal M}^{l}_{k}\vert H(6)^{jk}_{l}\vert {\cal B}_{i'i''}\rangle \epsilon ^{ii'i''}
+a_t(\overline{15})_{1}\langle {\cal F}^{k}_{l} {\cal M}^{l}_{j}\vert H(\overline{15})^{ij}_{k}\vert {\cal B}_{i'i''}\rangle \epsilon ^{ii'i''}\nonumber\\
&+&a_t(\overline{15})_{2} \langle {\cal F}^{l}_{j}{\cal M}^{k}_{l}\vert H(\overline{15})^{ij}_{k}\vert {\cal B}_{i'i''}\rangle \epsilon ^{ii'i''}
+b_t(\overline{15})_{1}\langle  {\cal F}^{l}_{k} {\cal M}^{i}_{j}\vert H(\overline{15})^{jk}_{l}\vert {\cal B}_{i'i''}\rangle \epsilon ^{ii'i''}\nonumber\\
&+&b_t(\overline{15})_{2}\langle {\cal F}^{i}_{j}{\cal M}^{l}_{k}\vert H(\overline{15})^{jk}_{l}\vert {\cal B}_{i'i''}\rangle \epsilon ^{ii'i''}\nonumber\\
&+&c_t(\overline{3})\langle {\cal M}^{i}_{j}{\cal F}^{i'}_{j'} \vert H(\overline{3})^{i''}  \vert {\cal B}_{jj'}\rangle \epsilon_{ii'i''}
+d_t(\overline{3})_{1}\langle {\cal M}^{i}_{j}{\cal F}^{i'}_{j'}\vert H(\overline{3})^{j}\vert {\cal B}_{i''j'}\rangle\epsilon_{ii'i''}\nonumber\\
&+&d_t(\overline{3})_{2}\langle {\cal F}^{i}_{j}{\cal M}^{i'}_{j'}\vert H(\overline{3})^{j}\vert {\cal B}_{i''j'}\rangle\epsilon_{ii'i''}
+e_t(\overline{3})_{1} \langle {\cal M}^{i}_{j'}{\cal F}^{i'}_{j}\vert H(\overline{3})^{j}\vert {\cal B}_{i''j'}\rangle\epsilon_{ii'i''}\nonumber\\
&+&e_t(\overline{3})_{2}\langle {\cal F}^{i}_{j'}{\cal M}^{i'}_{j}\vert H(\overline{3})^{j}\vert {\cal B}_{i''j'} \rangle \epsilon_{ii'i''}
+c_t(6)\langle {\cal M}^{i}_{j}{\cal F}^{i'}_{j'}\vert H(6)^{jj'}_{k}\vert {\cal B}_{i''k}\rangle\epsilon_{ii'i''}\nonumber\\
&+&d_t(6)_{1}\langle {\cal M}^{i}_{j}{\cal F}^{i'}_{j'}\vert H(6)^{i''j}_{k}\vert {\cal B}_{j'k}\rangle\epsilon_{ii'i''}
+d_t(6)_{2} \langle {\cal F}^{i}_{j}{\cal M}^{i'}_{j'}\vert H(6)^{i''j}_{k}\vert {\cal B}_{j'k}\rangle\epsilon_{ii'i''}\nonumber\\
&+&e_t(6)_{1} \langle {\cal M}^{i}_{j}{\cal F}^{i'}_{j'}\vert H(6)^{i''j'}_{k}\vert {\cal B}_{jk}\rangle\epsilon_{ii'i''}+
e_t(6)_{2}\langle {\cal F}^{i}_{j}{\cal M}^{i'}_{j'}\vert H(6)^{i''j'}_{k}\vert {\cal B}_{jk}\rangle\epsilon_{ii'i''}\nonumber\\
&+&f_t(6)\langle {\cal M}^{i}_{j}{\cal F}^{k}_{j'}\vert H(6)^{i'i''}_{k}\vert {\cal B}_{jj'}\rangle\epsilon_{ii'i''}+
g_t(6) \langle {\cal M}^{k}_{j}{\cal F}^{i}_{j'}\vert H(6)^{i'i''}_{k}\vert {\cal B}_{jj'}\rangle\epsilon_{ii'i''}\nonumber\\
&+&m_t(6) \langle {\cal M}^{k}_{j}{\cal F}^{j}_{k}\vert H(6)^{ii'}_{l}\vert {\cal B}_{i''l}\rangle\epsilon_{ii'i''}+
n_t(6)_{1}  \langle {\cal M}^{k}_{j}{\cal F}^{j}_{l}\vert H(6)^{ii'}_{k}\vert {\cal B}_{i''l}\rangle\epsilon_{ii'i''}\nonumber\\
&+&n_t(6)_{2}\langle  {\cal F}^{k}_{j}{\cal M}^{j}_{l}\vert H(6)^{ii'}_{k}\vert {\cal B}_{i''l}\rangle\epsilon_{ii'i''}
+c_t(\overline{15})\langle  {\cal M}^{i}_{j}{\cal F}^{i'}_{j'}\vert H(\overline{15})^{jj'}_{k}\vert {\cal B}_{i''k}\rangle\epsilon_{ii'i''}\nonumber\\
&+&d_t(\overline{15})_{1}\langle  {\cal M}^{i}_{j}{\cal F}^{i'}_{j'}\vert H(\overline{15})^{i''j}_{k}\vert {\cal B}_{j'k}\rangle\epsilon_{ii'i''}
+d_t(\overline{15})_{2} \langle {\cal F}^{i}_{j}{\cal M}^{i'}_{j'}\vert H(\overline{15})^{i''j}_{k}\vert {\cal B}_{j'k}\rangle\epsilon_{ii'i''}\nonumber\\
&+&e_t(\overline{15})_{1}\langle {\cal M}^{i}_{j}{\cal F}^{i'}_{j'}\vert H(\overline{15})^{i''j'}_{k}\vert {\cal B}_{jk}\rangle\epsilon_{ii'i''}+
e_t(\overline{15})_{2}\langle {\cal F}^{i}_{j}{\cal M}^{i'}_{j'}\vert H(\overline{15})^{i''j'}_{k}\vert {\cal B}_{jk}\rangle\epsilon_{ii'i''}\;
\end{eqnarray}
and
\begin{eqnarray}
T_{s}(q)_0&=&a_s(\overline{3})\langle {\cal M}^{i}_{j}{\cal F}^{i'}_{j'} \vert H(\overline{3})^{i''}  \vert {\cal C}_{jj'}\rangle \epsilon_{ii'i''}
+b_s(\overline{3})_{1}\langle {\cal M}^{i}_{j}{\cal F}^{i'}_{j'}\vert H(\overline{3})^{j}\vert {\cal C}_{i''j'}\rangle\epsilon_{ii'i''}\nonumber\\
&+&b_s(\overline{3})_{2}\langle {\cal F}^{i}_{j}{\cal M}^{i'}_{j'}\vert H(\overline{3})^{j}\vert {\cal C}_{i''j'}\rangle\epsilon_{ii'i''}
+c_s(\overline{3})_{1} \langle {\cal M}^{i}_{j'}{\cal F}^{i'}_{j}\vert H(\overline{3})^{j}\vert {\cal C}_{i''j'}\rangle\epsilon_{ii'i''}\nonumber\\
&+&c_s(\overline{3})_{2}\langle {\cal F}^{i}_{j'}{\cal M}^{i'}_{j}\vert H(\overline{3})^{j}\vert {\cal C}_{i''j'} \rangle \epsilon_{ii'i''}
+a_s(6)\langle {\cal M}^{i}_{j}{\cal F}^{i'}_{j'}\vert H(6)^{jj'}_{k}\vert {\cal C}_{i''k}\rangle\epsilon_{ii'i''}\nonumber\\
&+&b_s(6)_{1}\langle {\cal M}^{i}_{j}{\cal F}^{i'}_{j'}\vert H(6)^{i''j}_{k}\vert {\cal C}_{j'k}\rangle\epsilon_{ii'i''}
+b_s(6)_{2} \langle {\cal F}^{i}_{j}{\cal M}^{i'}_{j'}\vert H(6)^{i''j}_{k}\vert {\cal C}_{j'k}\rangle\epsilon_{ii'i''}\nonumber\\
&+&c_s(6)_{1} \langle {\cal M}^{i}_{j}{\cal F}^{i'}_{j'}\vert H(6)^{i''j'}_{k}\vert {\cal C}_{jk}\rangle\epsilon_{ii'i''}+
c_s(6)_{2}\langle {\cal F}^{i}_{j}{\cal M}^{i'}_{j'}\vert H(6)^{i''j'}_{k}\vert {\cal C}_{jk}\rangle\epsilon_{ii'i''}\nonumber\\
&+&d_s(6)\langle {\cal M}^{i}_{j}{\cal F}^{k}_{j'}\vert H(6)^{i'i''}_{k}\vert {\cal C}_{jj'}\rangle\epsilon_{ii'i''}+
e_s(6) \langle {\cal M}^{k}_{j}{\cal F}^{i}_{j'}\vert H(6)^{i'i''}_{k}\vert {\cal C}_{jj'}\rangle\epsilon_{ii'i''}\nonumber\\
&+&f_s(6) \langle {\cal M}^{k}_{j}{\cal F}^{j}_{k}\vert H(6)^{ii'}_{l}\vert {\cal C}_{i''l}\rangle\epsilon_{ii'i''}+
g_s(6)_{1}  \langle {\cal M}^{k}_{j}{\cal F}^{j}_{l}\vert H(6)^{ii'}_{k}\vert {\cal C}_{i''l}\rangle\epsilon_{ii'i''}\nonumber\\
&+&g_s(6)_{2}\langle  {\cal F}^{k}_{j}{\cal M}^{j}_{l}\vert H(6)^{ii'}_{k}\vert {\cal C}_{i''l}\rangle\epsilon_{ii'i''}
+a_s(\overline{15})\langle  {\cal M}^{i}_{j}{\cal F}^{i'}_{j'}\vert H(\overline{15})^{jj'}_{k}\vert {\cal C}_{i''k}\rangle\epsilon_{ii'i''}\nonumber\\
&+&b_s(\overline{15})_{1}\langle  {\cal M}^{i}_{j}{\cal F}^{i'}_{j'}\vert H(\overline{15})^{i''j}_{k}\vert {\cal C}_{j'k}\rangle\epsilon_{ii'i''}
+b_s(\overline{15})_{2} \langle {\cal F}^{i}_{j}{\cal M}^{i'}_{j'}\vert H(\overline{15})^{i''j}_{k}\vert {\cal C}_{j'k}\rangle\epsilon_{ii'i''}\nonumber\\
&+&c_s(\overline{15})_{1}\langle {\cal M}^{i}_{j}{\cal F}^{i'}_{j'}\vert H(\overline{15})^{i''j'}_{k}\vert {\cal C}_{jk}\rangle\epsilon_{ii'i''}+
c_s(\overline{15})_{2}\langle {\cal F}^{i}_{j}{\cal M}^{i'}_{j'}\vert H(\overline{15})^{i''j'}_{k}\vert {\cal C}_{jk}\rangle\epsilon_{ii'i''}\;.
\end{eqnarray}
For the invariant amplitudes above, although some of them are written with the same symble for both
anti-triplet and sextet, it should be understood that they are actually different ones.
The penguin amplitudes $P$ have the same structure and can be obtained to replace the expressions for $T$ by $P$. Also the P-wave amplitudes can be similarly constructed.

Expanding the above amplitudes, one can obtain the individual decay amplitude. Due to mixing between $\eta_8$ and $\eta_1$, the decay modes with $\eta_8$ in the final sates is not as clean as those with $\pi$ and $K$ in the final state to study. We will not consider processes involve $\eta_8$ further. We find the $U$-spin related amplitudes ($\Delta S=0$ and $\Delta S=-1$) for anti-triplet satisfy the following relations
\begin{eqnarray}
&& T_t(\Xi^{-}_{b} \to K^{-} n)= T_t(\Xi^{-}_{b} \to \pi^{-} \Xi^{0})\;, \;\;\;\;\;\;\;\;
 T_t(\Xi^{0}_{b} \to \bar{K}^{0} n)=-T_t(\Lambda^{0}_{b} \to K^{0} \Xi^{0})\;, \nonumber\\
&& T_t(\Xi^{-}_{b} \to K^{0} \Xi^{-})=T_t(\Xi^{-}_{b} \to \bar{K}^{0} \Sigma^{-})\;, \;\;\;\;\;
 T_t(\Xi^{0}_{b} \to K^{0} \Xi^{0})=-T_t(\Lambda^{0}_{b} \to \bar{K}^{0} n)\;, \nonumber\\
&& T_t(\Xi^{0}_{b} \to \pi^{-} \Sigma^{+})=-T_t(\Lambda^{0}_{b} \to K^{-} p)\;, \;\;\;\;\;\;
 T_t(\Lambda^{0}_{b} \to \pi^{-} p)=-T_t(\Xi^{0}_{b}\to K^{-} \Sigma^{+})\;,\nonumber\\
&& T_t(\Xi^{0}_{b} \to \pi^{+} \Sigma^{-})=-T_t(\Lambda^{0}_{b} \to K^{+} \Xi^{-})\;,\;\;\;
  T_t(\Lambda^{0}_{b} \to K^{+} \Sigma^{-})=-T_t(\Xi^{0}_{b} \to \pi^{+} \Xi^{-})\;,\nonumber\\
&& T_t(\Xi^{0}_{b} \to K^{-} p)=-T_t(\Lambda^{0}_{b} \to \pi^{-} \Sigma^{+})\;,\;\;\;\;\;\;
  T_t(\Xi^{0}_{b} \to K^{+} \Xi^{-})=-T_t(\Lambda^0_{b} \to \pi^{+} \Sigma^{-})\;. \label{r1}
\end{eqnarray}
While the $U$-spin related amplitudes ($\Delta S=0$ and $\Delta S=-1$) for sextet satisfy
\begin{eqnarray}
&& T_s(\Sigma^{+}_{b} \to n \pi^{+})=-T_s(\Sigma^{+}_{b} \to \Xi^{0} K^{+})\;,\;\;\;\;
T_s(\Sigma^{+}_{b} \to \Sigma^{+} K^{0})=-T_s(\Sigma^{+}_{b} \to p \bar{K}^{0})\;,\nonumber\\
&& T_s(\Sigma^{-}_{b} \to n \pi^{-})=-T_s(\Omega^{-}_{b} \to \Xi^{0} K^{-})\;,\;\;\;\;
 T_s(\Sigma^{-}_{b} \to \Sigma^{-} K^{0})=-T_s(\Omega^{-}_{b} \to \Xi^{-} \bar{K}^{0})\;,\nonumber\\
&& T_s(\Omega^{-}_{b} \to \Xi^{0} \pi^{-})=-T_s(\Sigma^{-}_{b} \to n K^{-})\;,\;\;\;\;
 T_s(\Omega^{-}_{b} \to \Sigma^{-} \bar{K}^{0})=-T_s(\Sigma^{-}_{b} \to \Xi^{-} K^{0})\;,\nonumber\\
&& T_s(\Sigma^{0}_{b} \to \Sigma^{-} K^{+})=-T_s(\Xi'^{0}_{b} \to \Xi^{-} \pi^{+})\;,\;\;
 T_s(\Sigma^{0}_{b} \to p \pi^{-})=-T_s(\Xi'^{0}_{b} \to \Sigma^{+} K^{-})\;,\nonumber\\
&&  T_s(\Xi'^{0}_{b} \to \Xi^{-} K^{+})=-T_s(\Sigma^{0}_{b} \to \Sigma^{-} \pi^{+})\;,\;\;
 T_s(\Xi'^{0}_{b} \to \Sigma^{-} \pi^{+})=-T_s(\Sigma^{0}_{b} \to \Xi^{-} K^{+})\;,\nonumber\\
&& T_s(\Xi'^{0}_{b} \to p K^{-})=-T_s(\Sigma^{0}_{b} \to \Sigma^{+} \pi^{-})\;,\;\;\;\;\;
  T_s(\Xi'^{0}_{b} \to \Sigma^{+} \pi^{-})=-T_s(\Sigma^{0}_{b} \to p K^{-})\;,\nonumber\\
&& T_s(\Xi'^{0}_{b} \to \Xi^{0} K^{0})=-T_s(\Sigma^{0}_{b} \to n \bar{K}^{0})\;,\;\;\;\;\;\;
 T_s(\Xi'^{0}_{b} \to n \bar{K}^{0})=-T_s(\Sigma^{0}_{b} \to \Xi^{0} K^{0})\;,\nonumber\\
&& T_s(\Xi'^{-}_{b} \to n K^{-})=-T_s(\Xi'^{-}_{b} \to \Xi^{0} \pi^{-})\;,\;\;\;
 T_s(\Xi'^{-}_{b} \to \Xi^{-} K^{0})=-T_s(\Xi'^{-}_{b} \to \Sigma^{-} \bar{K}^{0}). \label{r2}
\end{eqnarray}

\noindent
{\bf Results and Discussions}

For the processes in each of the pairs in eqs.(\ref{r1}) and (\ref{r2}), there are relations in the form given by eqs.(\ref{poasy}) and (\ref{raasy}). Some of them may be able to be tested at the LHCb. Experimentally it is difficult to measure the neutron in the final state. It is therefore not practical to carry out tests using decay modes with neutron in the final states. Also for relations associated with polarization asymmetries, one needs to measure the polarizations of initial or final baryon polarization. It is difficult to measure the initial baryon polarization at the LHC because the proton beams are not polarized. Information on  polarization can be extracted by the secondary decay of the final baryon. Therefore the final baryon which does not decay will not be useful for testing the relations associated with polarization asymmetries.

Relations associated with CP violating rate asymmetries for anti-triplet b-baryon decays have been studied recently. Practical tests for the relation
\begin{eqnarray}
&&{A_{CP}({\cal B}_a \to {\cal M} {\cal F} )_{\Delta S = 0}\over A_{CP}({\cal B}_b \to  {\cal M} {\cal F})_{\Delta S = -1}}
=
- {Br({\cal B}_b \to {\cal M} {\cal F} )_{\Delta S = -1}\over Br({\cal B}_a \to  {\cal M} {\cal F})_{\Delta S = 0}}\cdot
{{\tau_{{\cal B}_a} } \over  {\tau_{{\cal B}_b }} }\;,
\end{eqnarray}
can be carried out using the following pairs
\begin{eqnarray}
&&(\Xi^{-}_{b} \to K^{0} \Xi^{-},\;\;\Xi^{-}_{b} \to \bar{K}^{0} \Sigma^{-})\;, \nonumber\\
&&(\Xi^{0}_{b} \to \pi^{-} \Sigma^{+},\;\;\Lambda^{0}_{b} \to K^{-} p)\;, \;\;\;\;\;\;
 (\Lambda^{0}_{b} \to \pi^{-} p,\;\;\Xi^{0}_{b}\to K^{-} \Sigma^{+})\;,\nonumber\\
&&(\Xi^{0}_{b} \to \pi^{+} \Sigma^{-},\;\;\Lambda^{0}_{b} \to K^{+} \Xi^{-})\;,\;\;\;
(\Lambda^{0}_{b} \to K^{+} \Sigma^{-},\;\;\;\;\Xi^{0}_{b} \to \pi^{+} \Xi^{-})\;,\nonumber\\
&&(\Xi^{0}_{b} \to K^{-} p,\;\;\Lambda^{0}_{b} \to \pi^{-} \Sigma^{+})\;,\;\;\;\;\;\;
  (\Xi^{0}_{b} \to K^{+} \Xi^{-},\;\;\;\Lambda^0_{b} \to \pi^{+} \Sigma^{-})\;. \
\end{eqnarray}

For polarization asymmetry measurement, the final baryon should decay into other particles to provide polarization information. In the above, three of processes have the stable proton in the final state, therefore these decay modes are not useful for testing relations for polarization asymmetry. Only the following four pairs
\begin{eqnarray}
&&(\Xi^{-}_{b} \to K^{0} \Xi^{-},\;\;\Xi^{-}_{b} \to \bar{K}^{0} \Sigma^{-})\;, \;\;(\Xi^{0}_{b} \to \pi^{+} \Sigma^{-},\;\;\Lambda^{0}_{b} \to K^{+} \Xi^{-})\;,\nonumber\\
&&(\Lambda^{0}_{b} \to K^{+} \Sigma^{-},\;\;\Xi^{0}_{b} \to \pi^{+} \Xi^{-})\;,\;\;\;
  (\Xi^{0}_{b} \to K^{+} \Xi^{-},\;\;\Lambda^0_{b} \to \pi^{+} \Sigma^{-})\;. \label{rateasy}
\end{eqnarray}
may be useful in testing polarization asymmetry relation,
\begin{eqnarray}{{\cal A}_\alpha ({\cal B}_a \to {\cal M} {\cal F} )_{\Delta S = 0}\over {\cal A}_{\alpha}({\cal B}_b \to  {\cal M} {\cal F})_{\Delta S = -1}}
=
- {Br({\cal B}_b \to {\cal M} {\cal F} )_{\Delta S = -1}\over Br({\cal B}_a \to  {\cal M} {\cal F})_{\Delta S = 0}}\cdot
{{\tau_{{\cal B}_a} } \over  {\tau_{{\cal B}_b }} }\;.\label{polasy}
\end{eqnarray}

The initial particles in the pair $\Xi^{-}_{b} \to K^{0} \Xi^{-}$ and $\Xi^{-}_{b} \to \bar{K}^{0} \Sigma^{-}$ decays are the same making the above corresponding equation simpler than other pairs above since no lifetime information needed. These may be the best modes to carry out analysis for processes involve the anti-triplet b-baryons experimentally.

For sextet, it is possible to test the rate asymmetry relation of the type in eq.(\ref{rateasy}) using the pairs below
\begin{eqnarray}
&&
(\Sigma^{+}_{b} \to \Sigma^{+} K^{0}\;,\;\;\Sigma^{+}_{b} \to p \bar{K}^{0})\;,\;\;\;\;\;(\Sigma^{-}_{b} \to \Sigma^{-} K^{0}\;,\;\;\Omega^{-}_{b} \to \Xi^{-} \bar{K}^{0})\;,\nonumber\\
&&
 (\Omega^{-}_{b} \to \Sigma^{-} \bar{K}^{0}\;,\;\;\Sigma^{-}_{b} \to \Xi^{-} K^{0})\;,\;\;(\Sigma^{0}_{b} \to \Sigma^{-} K^{+}\;,\;\;\Xi'^{0}_{b} \to \Xi^{-} \pi^{+})\;,\nonumber\\
 &&(\Sigma^{0}_{b} \to p \pi^{-}\;,\;\;\Xi'^{0}_{b} \to \Sigma^{+} K^{-})\;,\;\;\;\;\;\;(\Xi'^{0}_{b} \to \Xi^{-} K^{+}\;,\;\;\;\;\Sigma^{0}_{b} \to \Sigma^{-} \pi^{+})\;,\nonumber\\
&&(\Xi'^{0}_{b} \to \Sigma^{-} \pi^{+}\;,\;\;\Sigma^{0}_{b} \to \Xi^{-} K^{+})\;,\;\;\;(\Xi'^{0}_{b} \to p K^{-}\;,\;\;\Sigma^{0}_{b} \to \Sigma^{+} \pi^{-})\;,\nonumber\\
&&(\Xi'^{0}_{b} \to \Sigma^{+} \pi^{-}\;,\;\;\Sigma^{0}_{b} \to p K^{-})\;,\;\;\;\;\;\;(\Xi'^{-}_{b} \to \Xi^{-} K^{0}\;,\;\;\Xi'^{-}_{b} \to \Sigma^{-} \bar{K}^{0}).
\end{eqnarray}

For the polarization asymmetry relation test, one again has to remove the ones with stable proton in the final state. Therefore the following pairs are good ones for the polarization asymmetry relation test,
\begin{eqnarray}
&&
(\Sigma^{-}_{b} \to \Sigma^{-} K^{0}\;,\;\;\Omega^{-}_{b} \to \Xi^{-} \bar{K}^{0})\;,\;\;
(\Omega^{-}_{b} \to \Sigma^{-} \bar{K}^{0}\;,\;\;\Sigma^{-}_{b} \to \Xi^{-} K^{0})\;,\nonumber\\
&&(\Sigma^{0}_{b} \to \Sigma^{-} K^{+}\;,\;\;\Xi'^{0}_{b} \to \Xi^{-} \pi^{+})\;,\;\;\;(\Xi'^{0}_{b} \to \Xi^{-} K^{+}\;,\;\;\Sigma^{0}_{b} \to \Sigma^{-} \pi^{+})\;,\nonumber\\
&&(\Xi'^{0}_{b} \to \Sigma^{-} \pi^{+}\;,\;\;\Sigma^{0}_{b} \to \Xi^{-} K^{+})\;,\;\;\;(\Xi'^{-}_{b} \to \Xi^{-} K^{0}\;,\;\;\Xi'^{-}_{b} \to \Sigma^{-} \bar{K}^{0}).
\end{eqnarray}

For sextet, the best pair to carry out analysis may be the last pair above, $(\Xi'^{-}_{b} \to \Xi^{-} K^{0}\;,\;\;\Xi'^{-}_{b} \to \Sigma^{-} \bar{K}^{0})$ because that the initial particles are the same.

To actually test the relations discussed here, it is desirable that the CP violating rate and polarization asymmetries be reasonablely large. Some theoretical estimates for some of the decays have been carried out~\cite{model}. For example,
for the rate asymmetries for $\Lambda_b \to p \pi^-,\;\; p K^-$ are of order a few percent~\cite{model} which  are reasonablly sizeable for testing the relations studied here.

$SU(3)$ breaking effects are known to exist to some degree. In Kaon and Hyperon decays, the breaking effects are at the order of 20 to 30 percent. One may expect that the breaking effects are at a similar level and therefore the relations studied hold also at that level. However, the CP violating rate asymmetry relations similar to eq.(\ref{raasy}) for
the pair $\bar B^0_s \to K^+ \pi^-$ and $\bar B^0\to K^- \pi^+$ seems to hold at a better level. If one defines a measure of $SU(3)$ breaking by
$r_c =-[ A_{CP}(\bar B^0_s \to K^+ \pi^-)/ A_{CP}(\bar B^0\to K^- \pi^+)]/
[Br(\bar B^0 \to K^- \pi^+) \tau_{\bar B^0_s} / Br(\bar B^0_s \to K^+ \pi^-)\tau_{\bar B^0} ]$. In the $SU(3)$ limit, $r_c = 1$. Experimentally\cite{pdg,rateasymmetry, he4} $r_c = 0.96\pm 0.19$. The central value is about 5\% away from 1. The 1$\sigma$ level error bar is about 20\%. This is an indication that $SU(3)$ may work better in systems with a b quark than that for Kaon and Hyperon systems. Whether this is an accidental or $SU(3)$ works better for B decays needs to be understood. It is therefore important to test relations discussed here experimentally. Experimental data obtained will provide crucial information to understand the dynamics of b-hadron decays. We eagerly urge our experimental colleagues to carry out related test discussed here.

\begin{acknowledgments}

The work was supported in part by MOE Academic Excellent Program (Grant No: 102R891505) and MOST of ROC, and in part by NSFC(Grant No:11175115) and Shanghai Science and Technology Commission (Grant No: 11DZ2260700) of PRC.

\end{acknowledgments}

\end{document}